\documentclass{moriond}

\usepackage{graphicx}
\usepackage{amssymb} 
\usepackage{amsthm} 
\usepackage{lineno}
\usepackage{hyperref}
\usepackage{sidecap}

\def\be{\begin{equation}}
\def\ee{\end{equation}}
\def\bea{\begin{eqnarray}}
\def\eea{\end{eqnarray}}

\newcommand{\sqrtsNN}{\sqrt{s_{\scriptscriptstyle \rm NN}}}

\newcommand{\GeV}{\mathrm{GeV}}

\newcommand{\gev}{\mathrm{GeV}}

\newcommand{\tev}{\mathrm{TeV}}

\newcommand{\PbPb}{\mbox{Pb--Pb}}
\newcommand{\pPb}{\mbox{p--Pb}}

\newcommand{\pt}{p_{\rm T}}
\newcommand{\pT}{p_{\rm T}}

\newcommand{\Dzero}{{\rm D^0}}

\newcommand{\Dstar}{{\rm D^{*+}}}
\newcommand{\Ds}{{\rm D_{s}^{+}}}

\newcommand{\Dplus}{{\rm D^+}}

\newcommand{\Jpsi}{{\rm J}/\psi}

\renewcommand{\PbPb}{\mbox{Pb--Pb}}

\newcommand{\Raa}{R_{\rm AA}}
\newcommand{\RpPb}{R_{\rm pPb}}
\newcommand{\RAA}{R_{\rm AA}}

\newcommand{\Photo}{}

\begin{document}
\vspace*{4cm}
\title{Open heavy-flavour and quarkonium production in Pb-Pb and p-Pb
  collisions measured by the ALICE detector at the LHC}

\author{D. Caffarri for the ALICE Collaboration}

\address{CERN, Geneva, Switzerland}


\maketitle\abstracts{
Open heavy-flavour and quarkonia measurements are important tools to
study the hot and dense partonic medium formed in ultra-relativistic heavy-ion
collisions. The modification of their production in those collisions,
with respect to the pp and $\pPb$ ones, can help in the characterization
of this medium. Quarkonia and open heavy-flavour production is
measured in ALICE in the three different collision systems, at mid-
and forward rapidity. A selection of
those results recently
obtained in $\PbPb$ and $\pPb$ collisions by the ALICE Collaboration is presented.}

\section{Introduction}
The Large Hadron Collider (LHC) allows the study of ultra-relativistic
collisions of heavy-ions, in particular Pb-Pb and p-Pb collisions.
The ALICE experiment~\cite{Abelev:2014ffa} was built to study in detail these interactions involving ions, in order to
characterize the deconfined, highly dense and hot state of nuclear
matter, known as the Quark-Gluon Plasma (QGP).

Pairs of charm and anti-charm quarks can be
produced in the scattering between two partons with very high momentum
transferred. At the LHC energy, those pairs are produced mainly via
gluon fusion at tree level process or via gluon splitting or flavour
excitation at higher order processes~\cite{Mangano:1991jk}. Charm
quarks that hadronise with
light quarks form open heavy-flavour hadrons; in case
heavy quarks pair among themselves, a quarkonium bound state
is produced. A non-perturbative
approach needs to be considered, when the relative velocity of the
quarks pair is similar to the quarks pair
mass~\cite{Kramer:2001hh}. For the quarkonium case, non-relativistic
QCD approaches, including colour-singlet and colour-octet fragmentation
processes were found to improve the agreement between data and calculations, albeit dominated by the still large theoretical uncertainties~\cite{Butenschoen:2011yh}.

The modification of the production yields in Pb-Pb collisions allow to
study how charm quarks interact with the medium and, in particular,
how they lose energy while passing through it. This energy loss can
occur via elastic scatterings of heavy quarks with other partons
of the medium or via inelastic processes that induce gluon
radiation. Theoretical calculations show that this energy loss
depends on the colour charge and the mass of the parton that traverses
the medium. Gluons should lose more energy than light quarks due to
their larger Casimir factor~\protect\cite{Baier:1996kr,Salgado:2003gb}. 
Heavy quarks are expected to lose less energy than
light quarks due to their larger mass that reduces the probability of
gluon emission~\protect\cite{Dokshitzer:1991fd,Dokshitzer:2001zm}. 
Quarkonia, instead, can escape from the medium only 
if their binding energy is larger than the colour screening potential
generated by the deconfined medium~\cite{MatsuiSatz}. Excited
quarkonia states are more likely to melt in the medium, because their
binding energy is smaller with respect to their correspondent ground states.
Due to the large number of $c\bar{c}$ pairs produced in the collisions at LHC
energies, quarkonia states could also be ``regenerated'' from
$c\bar{c}$ quarks produced in different hard scatterings~\cite{BraunMunzinger:2000px}.
Cold Nuclear
Matter (CNM) effects, as modification of the Parton
Distribution Functions in the nuclei~\cite{Eskola:2009uj}, gluon
radiation~\cite{Arleo:2010rb} or comover
interactions~\cite{Capella:2000zp} can modify the heavy-quark
production. These effects can be studied via p-Pb collisions
that were first delivered by the LHC in 2013. 

Open heavy-flavour particles produced at mid-rapidity ($|\eta|<0.9$) are
measured in ALICE~\cite{Abelev:2014ffa} by the full reconstruction of D-meson decay
topologies with displaced vertices and by measuring the spectra of
electrons from open heavy-flavour hadron decays. At forward rapidity ($-4<
\eta< -2.5$), their production is studied via muons coming from open
heavy-flavour hadron decays. Quarkonia production is measured at
mid (forward) rapidity via the di-electron (muon) decay channel; for
both cases quarkonia are reconstructed down to transverse momentum
$\pt$=0. 

Results will be presented in terms of the nuclear modification factor
$\Raa$ ($\RpPb$): the ratio of the spectra measured in $\PbPb$ ($\pPb$)
collisions, scaled by the number of binary nucleon-nucleon collisions,
divided by the one measured in pp collisions, as a function of transverse momentum ($\pt$) or rapidity ($y$).  
 
\section{Open heavy-flavour results}

\begin{figure}[t]
\begin{center}
\includegraphics[width=0.48\textwidth]{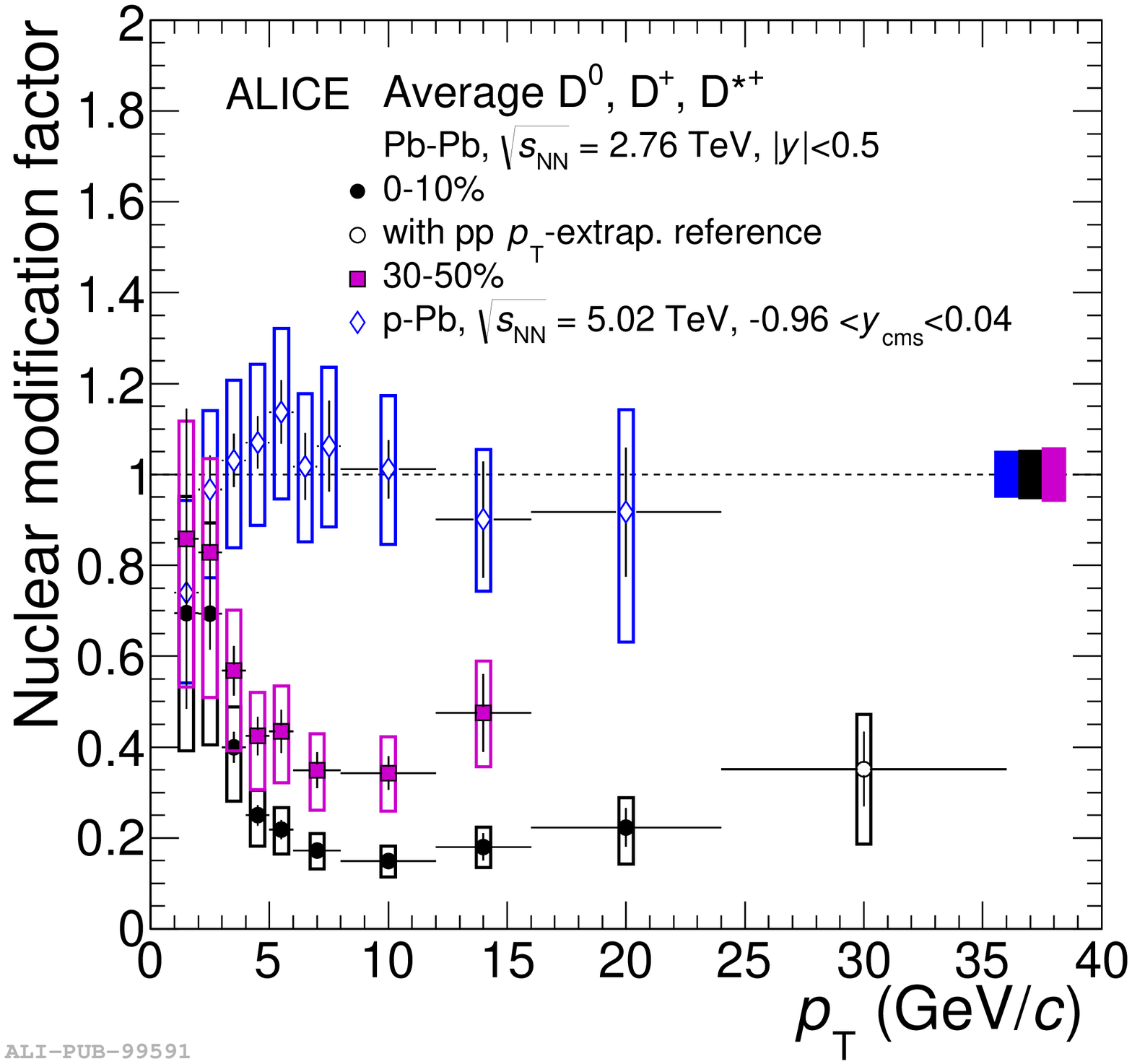}
\includegraphics[width=0.48\textwidth]{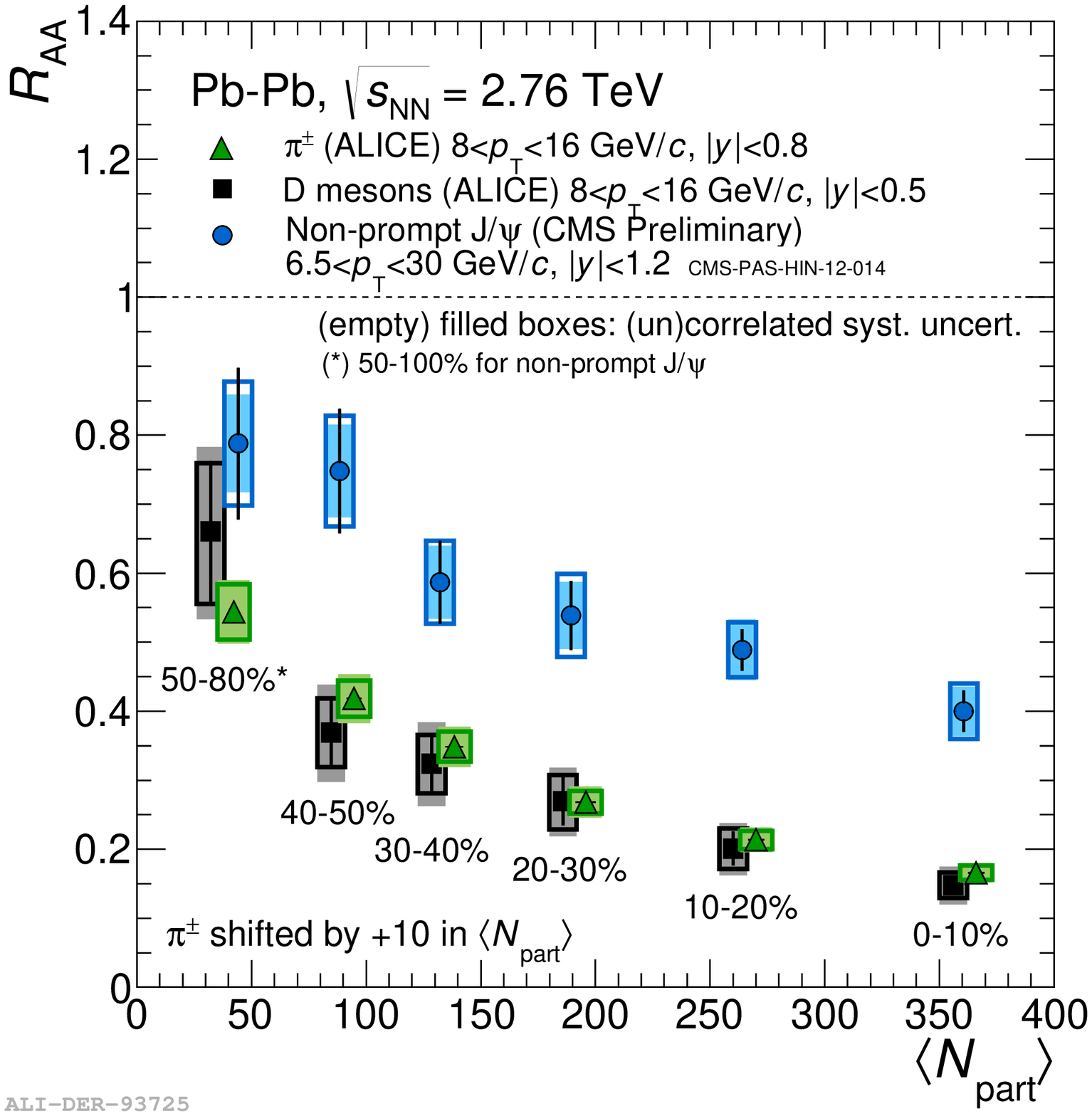}
\end{center}
\caption{Left: Prompt D-meson $\Raa$ (average of $\Dzero$, $\Dplus$ and 
$\Dstar$) as a function of $\pt$ in $\PbPb$ collisions at 
 $\sqrtsNN=2.76~\tev$ in the 0--10\% and 30--50\% centrality
 classes~\protect\cite{Adam:2015sza}. Prompt D-meson nuclear
 modification factor $\RpPb$ (average of $\Dzero$, $\Dplus$ and 
$\Dstar$) as a function of $\pt$ in $\pPb$ collisions at 
 $\sqrtsNN=5.02~\tev$~\protect\cite{Abelev:2014hha}. Right: Comparison
 of the average D-meson $\Raa$ in $8<\pt<16~\gev/c$ with charged pions
 in the same $\pt$ range ~\protect\cite{Adam:2015nna} and $\Jpsi$ from B
 decays in $6.5<\pt<30~\gev/c$.}
\label{fig:HFDmes}
\end{figure}

In order to study the interaction of charm quarks with the medium, the
D-meson nuclear modification factor in $\PbPb$ collisions has been
measured by ALICE as a function of transverse momentum as reported
in Fig.~\ref{fig:HFDmes}, left, for
two centrality classes (0-10\% and 30-50\%)~\cite{Adam:2015sza}.
A suppression of about a factor 5-6 at $\pt \sim 10~\GeV/c$ is observed for the most central
collisions. For the centrality class 30-50\% the suppression is reduced to a
factor about 3 in a similar momentum range. In the same figure also
the D-meson $\RpPb$ is shown and it is compatible with unity~\cite{Abelev:2014hha}. This
result confirms that the suppression observed in $\PbPb$ collisions
comes from an interaction of partons with the hot and dense
nuclear medium.
ALICE measured also the $\RpPb$ of leptons from open
heavy-flavour decays and no difference from unity
was found at backward, central and forward rapidity. Models that
include Cold Nuclear Matter effects are in good agreement with the
measurements~\cite{Adam:2015qda}.

The D-meson $\Raa$ was also been studied by ALICE as a function of the
centrality of the collisions: a larger suppression
is observed for central collisions than for peripheral ones. For the momentum range
$8<\pT<16~\GeV/c$ (Fig.~\ref{fig:HFDmes}, right)~\cite{Adam:2015nna}, the
D-meson results are compared to non-prompt $\Jpsi$, coming
from B-hadron decay, measured by
the CMS experiment~\cite{CMSJpsi2011} in an equivalent kinematic range and to charged
pions measured by ALICE. A larger
suppression is observed for D mesons than for non-prompt $\Jpsi$ while it is
similar for $\pi^{\pm}$ and D mesons. Theoretical calculations that
include a dependence on parton mass and colour
charge of the energy loss, can reproduce the results~\cite{Adam:2015nna}. 

The study of the $\Ds$-meson production in $\PbPb$ collisions is
sensitive to the strangeness enhancement observed in heavy-ion
collisions. ALICE performed this measurement for the first time in
central $\PbPb$ collisions~\cite{Adam:2015jda}. The results for strange and non-strange D
mesons are compatible within uncertainties and no conclusions can be
drawn from the current Run1 data.

\section{Quarkonia results} 

\begin{figure}[t]
\begin{center}
\includegraphics[width=0.48\textwidth]{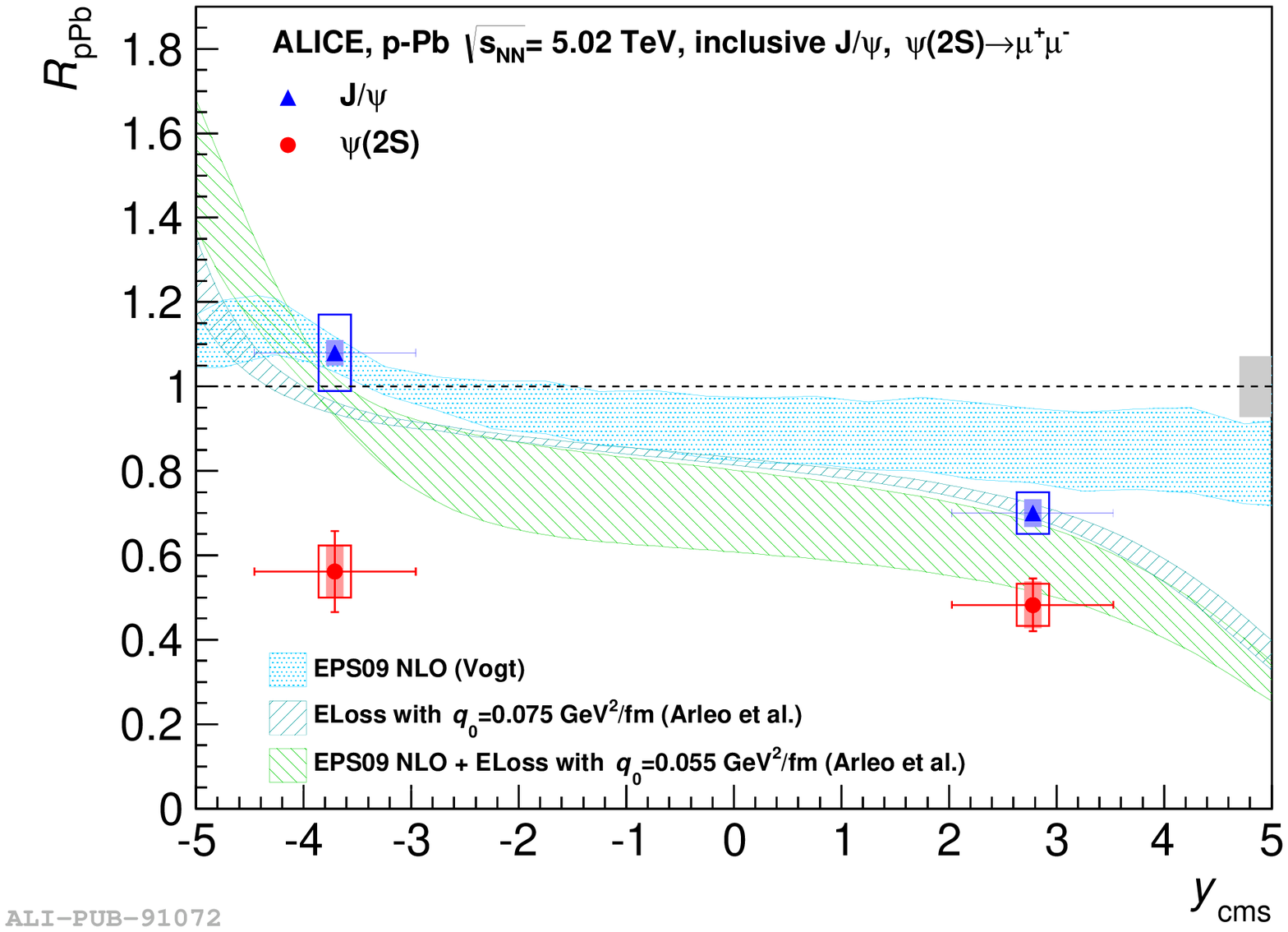}
\includegraphics[width=0.48\textwidth]{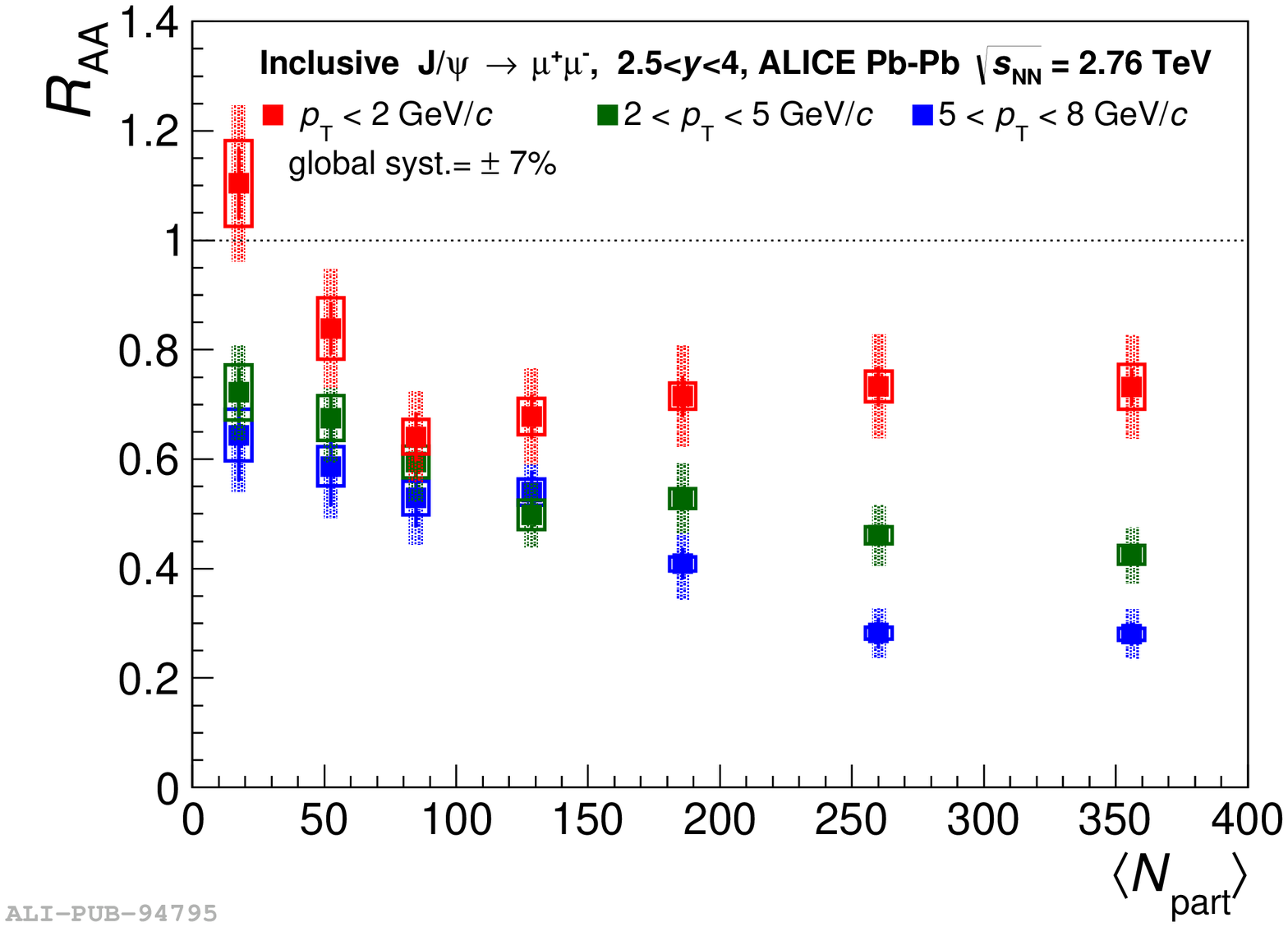}
\end{center}
\caption{Left: $\Jpsi$ and $\psi(2\rm{S})$ $\RpPb$ as a function of
  rapidity integrated over $\pt$~\protect\cite{Abelev:2014zpa}. Right $\Jpsi$ nuclear modification
  factor measured in $\PbPb$ collisions at  $\sqrtsNN=2.76~\tev$ as a
  function of centrality for three transverse momentum intervals~\protect \cite{Adam:2015isa}.}
\label{fig:DQonia}
\end{figure}

ALICE measured the $\RpPb$ of $\Jpsi$ in three different rapidity
regions: a suppression is observed in the central and forward region,
differently to what is measured at backward rapidity where no
suppression is reported~\cite{Adam:2015iga}.
The results are presented as a function of rapidity in
Fig.~\ref{fig:DQonia}, left, integrated over
$\pT$~\cite{Abelev:2014zpa}. The results show a good agreement with models that
include shadowing, coherent energy loss or Colour Glass Condensate
calculations~\cite{Adam:2015iga}. 

The measurement of the $\RpPb$ of the excited state $\psi(2\rm{S})$ has
also been performed by ALICE~\cite{Abelev:2014zpa}. Results are presented in
Fig.~\ref{fig:DQonia}, left. A similar suppression has been observed at
backward and forward rapidity, differently from what is observed for
the $\Jpsi$ and to what it would be expected, considering the
similar initial-state effects for the two charmonium states. 
Models that include shadowing or coherent energy loss cannot describe
the suppression observed at backward rapidity. ALICE performed also
more differential studies on the $\psi(2\rm{S})$ suppression and no significant
$\pT$ dependence is observed~\cite{Abelev:2014zpa}. A larger difference between the two
states, instead, is observed for central events, in the backward region~\cite{Adam:2016ohd}. In the
forward region, $\Jpsi$ and $\psi(2\rm{S})$ show very similar patterns as a
function of the multiplicity of the collision.
Models that include break up of the resonance due to comovers or
hadron gas in the final state reproduce the observed trend~\cite{Adam:2016ohd}. 

The $\RAA$ of $\Jpsi$ at forward rapidity in $\PbPb$ collisions is shown
as a function of centrality for three different momentum intervals in
Fig.~\ref{fig:DQonia}, right~\cite{Adam:2015isa}. $\Jpsi$ with $\pt > 2~\GeV/c$ show a larger suppression
in central collisions than in the peripheral ones. Low-$\pT$
$\Jpsi$, instead, show flatter trend as a function of centrality. 
The CNM effects inferred from the $\Jpsi$ suppression in $\pPb$
collisions at the forward rapidity region are not enough to explain
the suppression observed in $\PbPb$ collisions. These results are, indeed, in agreement with models that include melting of the
$\Jpsi$ in the medium due to the colour screening and $\Jpsi$ regeneration~\cite{Adam:2015isa}.

\begin{figure}[t]
\begin{center}
\includegraphics[width=0.48\textwidth]{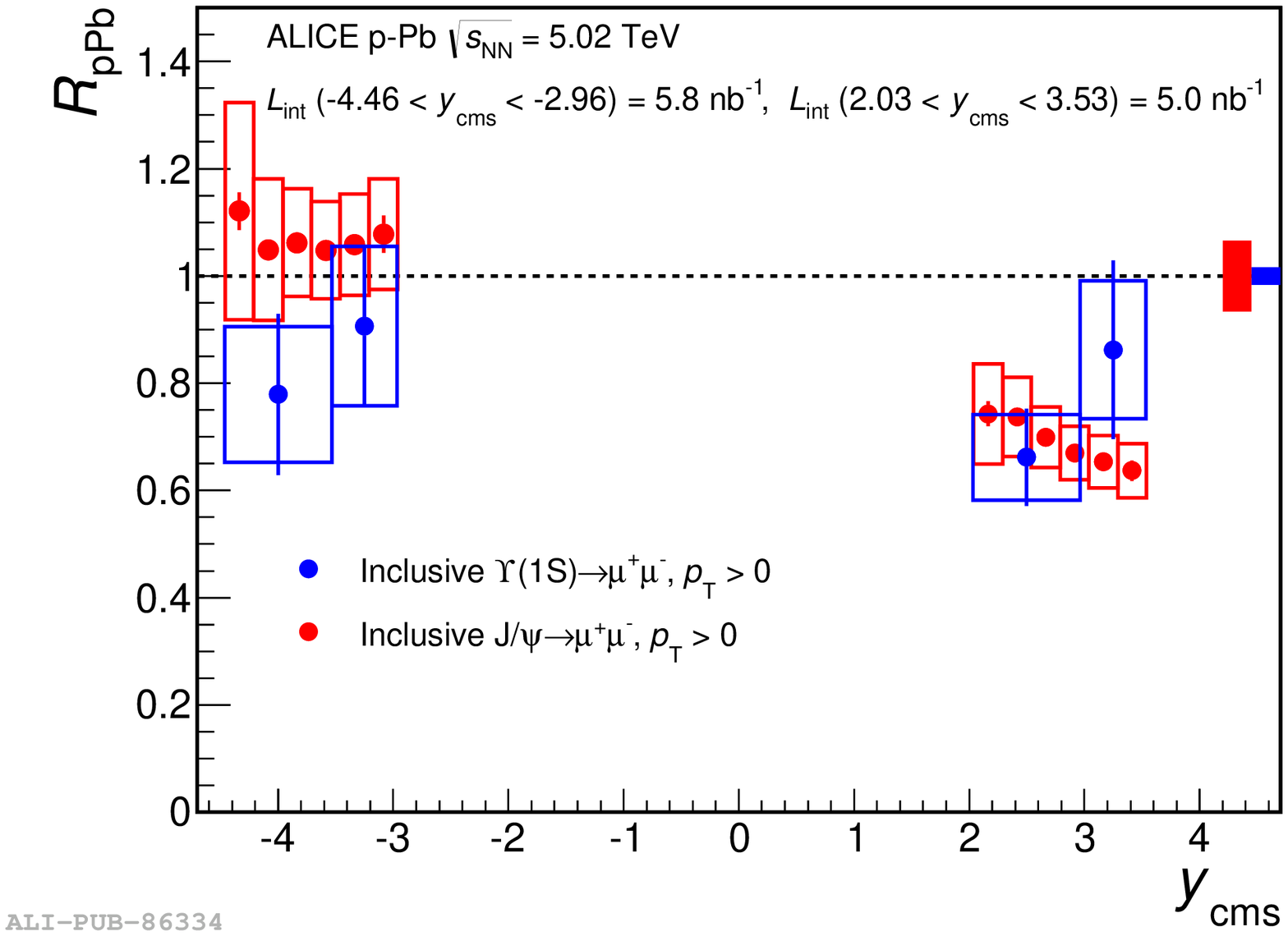}
\includegraphics[width=0.48\textwidth]{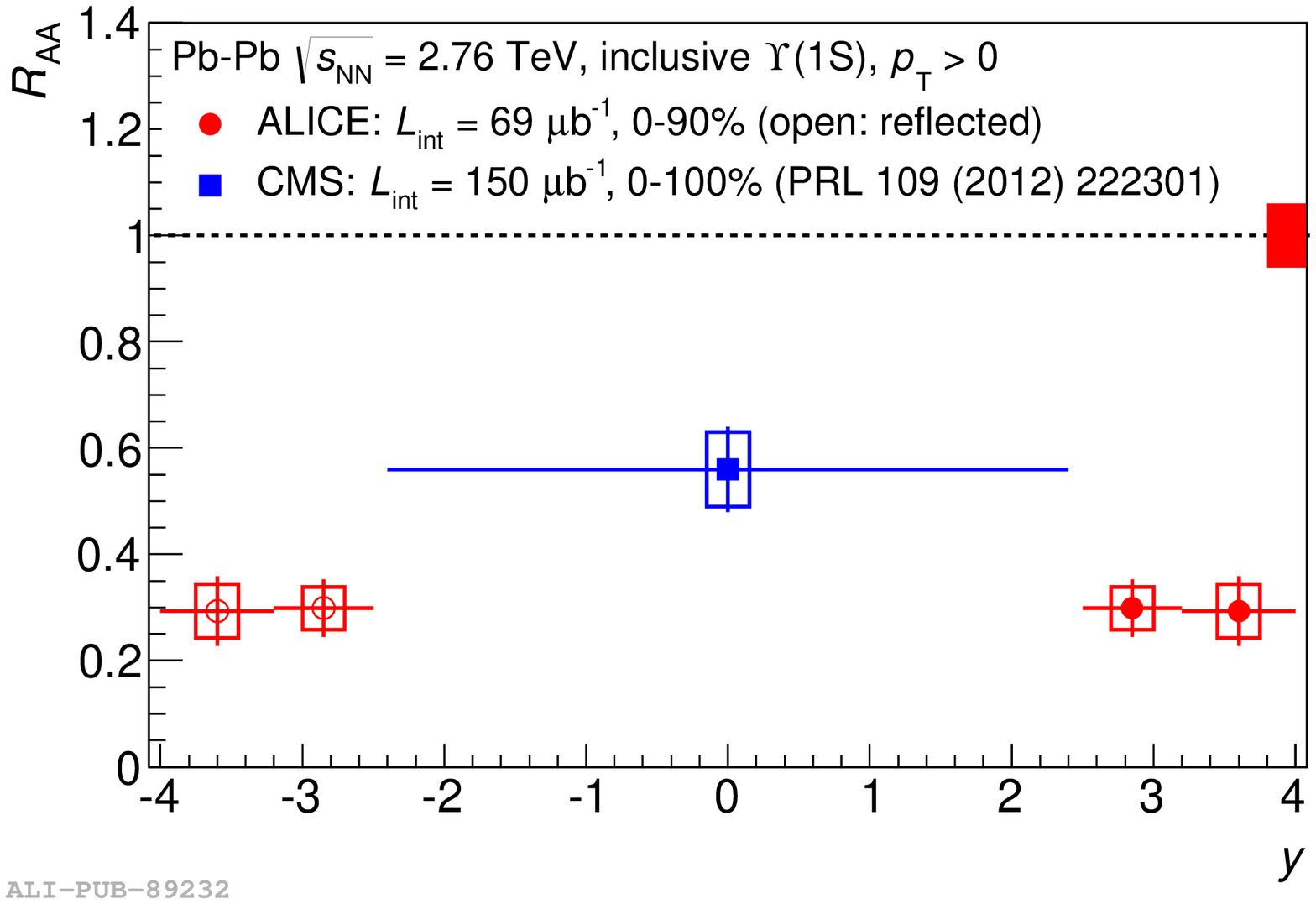}
\end{center}
\caption{Left: $\Upsilon(1\rm{S})$ $\RpPb$ as a function of
  rapidity integrated over $\pt$~\protect\cite{Abelev:2014oea}. Right: $\Upsilon(1\rm{S})$ nuclear modification
  factor measured by ALICE at forward rapidity~\protect\cite{Abelev:2014nua} compared with CMS
  $\Upsilon(1\rm{S})$ $\Raa$ at central
  rapidity.~\protect\cite{Chatrchyan:2012lxa}}
\label{fig:bottomonia}
\end{figure}

ALICE also published the results of the $\Upsilon(1\rm{S})$ $\RpPb$,
showing a similar value as the $\Jpsi$, in the
forward and backward rapidity regions, within uncertainties~\cite{Abelev:2014oea}
(Fig.~\ref{fig:bottomonia}, left). The results are in agreement with
theoretical calculations, though their current large uncertainties. 
In $\PbPb$ collisions, instead, a strong suppression of
$\Upsilon(1\rm{S})$ state is observed, as shown in Fig.~\ref{fig:bottomonia}, right. 
Comparing the ALICE and CMS results~\cite{Chatrchyan:2012lxa}, the suppression appears to be
larger at forward than at central rapidities~\cite{Abelev:2014nua}.

\end{document}